\documentclass{article}
\usepackage{arxiv}

\usepackage[hyphens]{url}  
\usepackage{graphicx} 
\usepackage[numbers, compress]{natbib}

\usepackage{amsmath}
\usepackage{amssymb}
\usepackage{xcolor}         
\usepackage{booktabs}
\usepackage{layouts}
\usepackage{multicol}

\definecolor{linkblue}{RGB}{50,100,170}
\usepackage{hyperref}
\hypersetup{
    pdfborder={0 0 0},
    colorlinks=true,
    urlcolor=linkblue,
    linkcolor=linkblue,
    citecolor=linkblue,
    bookmarks=true,
    bookmarksnumbered=true
}

\title{Quality Control in Weather Monitoring with Dynamic Linear Models}
\date{}
\author{%
    Joel Janek Dabrowski$^*$ \\
    Data61, CSIRO, \\
    Australia \\
    \texttt{Joel.Dabrowski@data61.csiro.au}
    \And
    Ashfaqur Rahman \\
    Data61, CSIRO, \\
    Australia \\
    \texttt{Ashfaqur.Rahman@data61.csiro.au}
    \And
    Ming Li \\
    Data61, CSIRO, \\
    Australia \\
    \texttt{Ming.Li@data61.csiro.au}
    \And
    Quanxi Shao \\
    Data61, CSIRO, \\
    Australia \\
    \texttt{Quanxi.Shao@data61.csiro.au}
    \And
    Shuvo Bakar \\
    Data61, CSIRO, \\
    Australia \\
    \texttt{Shuvo.Bakar@data61.csiro.au}
    \And
    Andrea Powell \\
    Data61, CSIRO, \\
    Australia \\
    \texttt{Andrea.Powell@data61.csiro.au}
    \And
    Brent Henderson \\
    Data61, CSIRO, \\
    Australia \\
    \texttt{Brent.Henderson@data61.csiro.au}
}

\newcommand{\h}{\boldsymbol{h}}
\renewcommand{\v}{\boldsymbol{v}}
\newcommand{\f}{\boldsymbol{f}}
\newcommand{\A}{\boldsymbol{A}}
\newcommand{\B}{\boldsymbol{B}}
\newcommand{\I}{\boldsymbol{I}}
\newcommand{\F}{\boldsymbol{F}}
\newcommand{\K}{\boldsymbol{K}}
\newcommand{\C}{\boldsymbol{C}}
\newcommand{\Sigmabf}{\boldsymbol{\Sigma}}
\newcommand{\epsilonbf}{\boldsymbol{\epsilon}}
\newcommand{\etabf}{\boldsymbol{\eta}}
\newcommand{\mubf}{\boldsymbol{\mu}}

\makeatletter
\def\blfootnote{\gdef\@thefnmark{}\@footnotetext}
\makeatother

\begin{document}

\maketitle

\begin{abstract}
    Decisions in agriculture are frequently based on weather. With an increase in the availability and affordability of off-the-shelf weather stations, farmers able to acquire localised weather information. However, with uncertainty in the sensor and installation quality, farmers are at risk of making poor decisions based on incorrect data. We present an automated approach to perform quality control on weather sensors. Our approach uses time-series modelling and data fusion with Bayesian principles to provide predictions with uncertainty quantification. These predictions and uncertainty are used to estimate the validity of a sensor observation. We test on temperature, wind, and humidity data and achieve error hit rates above 80\% and false negative rates below 11\%.\blfootnote{Published in ``The 2nd AAAI Workshop on AI for Agriculture and Food Systems'', 2023,\\ \url{https://openreview.net/forum?id=OP7Mq-z9SZ-}.}
\end{abstract}

\section{Introduction}

Making decisions based on weather is common practice in agriculture. Government agencies such as the Australian Bureau of Meteorology (BoM) install weather stations across Australia to monitor weather conditions. The data generated from these stations can be used for forecasting extreme heat, snow, frost, and heavy rainfall, which directly affect farmers. Long-term weather forecasts can also be used understand climatic changes and it's affects.

While the BoM weather stations provide sufficient geographical coverage for large-scale weather modelling, the coverage can be insufficient for localised weather modelling. For example, temperature, rainfall, and wind can be highly localised to small geographic regions. Furthermore, BoM weather stations are often located to provide coverage of urban regions, whereas farms are typically located in remote regions. Farmers thus often install off-the-shelf weather stations on their farms; which we refer to as Third Party Weather Stations (TPWS).


Unlike BoM weather stations, TPWS data may not be quality controlled, resulting in uncertainty on the reliability and correctness of the data. For example, low cost sensors may be inaccurate or the weather stations may be incorrectly installed. Incorrect data may cause a farmer take unnecessary costly actions or fail to take actions leading to disastrous consequences. 

We present an automated approach to detect unusual readings from TPWS data. The approach uses the Dynamic Linear Model (DLM) and the Kalman filter. It performs time series analysis and data fusion using Bayesian principles to provide predictions with uncertainty quantification. A key contribution is that we provide a means to automatically calibrate model. We show that this approach is robust in that the same model, parameters, and hyper-parameters are used across various weather stations and variables to provide accurate results with error hit rates above 80\% and false positive rates below 11\%. This work is especially relevant to those who seek a solution which is computationally efficient, easy to implement, and easy to maintain.

%
%

\section{Related Work}

Quality control for weather sensing has been an active research topic for several years \cite{leung2007effective,kenneth2007improved,mateo2008design} and a recent surge has come about in relation to the so-called ``citizen'' or ``crowdsourced'' weather stations \cite{beele2022quality,meier2017crowdsourcing,napoly2018development}. 
The common approaches are heuristic, statistical, or regression based. 
Heuristic approaches include thresholds, duplicate checks, and step-change tests. 
Statistical approaches compare samples to the mean and standard deviations of cyclic data and previous extrema. 
Regression approaches compare samples to neighbouring station interpolations, previous forecasts, and daily observations. 
Unlike the proposed approach, none of these approaches consider sub-daily dynamics of the variables in a time-series sense. The more recent approaches tend to take advantage of the high spatial concentration of ``crowdsourced'' weather stations in urban environments. High concentrations are however not likely in an agricultural setting. 

The DLM (also called the linear dynamic system, linear Gaussian state-space model, or Kalman filter) is known for its ability for state estimation amongst noise \cite{murphy2012machine}. It is also well equipped to perform outlier detection in time-series (e.g. \cite{ting2007learning,zhang2009study}). Furthermore, Kalman filters are well-known amongst the meteorological community (usually as Ensemble Kalman filters for NWP) \cite{roh2013observation,evensen2009data}. However, little work has been done in developing it for quality control of weather sensors. 

\section{Methodology}
\label{sec:methodology}


A DLM is employed to make predictions of a given weather variable over time. It fuses data from a TPWS and from a Numerical Weather Prediction (NWP) product, and also models the dynamics of the data over time. The Kalman filter is used to perform inference in the DLM by updating the weather variable predictions at each time given past data. A weather variable prediction is the form of a Gaussian distribution. If a given test observation is an outlier relative to the predicted Gaussian, it is classified as a suspect sample. 

The DLM is automatically calibrated using an initial set of data with length $T$. This calibration is based on a noise estimate of this initial dataset. 

%

%
%

\subsection{Dynamic Linear Model}

The DLM is a linear-Gaussian state-space model \cite{west2006bayesian,murphy2012machine}. It comprises a latent variable $\h_t \in \mathbb{R}^{H \times 1}$ and an observed variable $\v_t \in \mathbb{R}^{V \times 1}$. In this study, $\h_t$ represents modelled time-series components the particular weather variable of interest and $\v_t$ represents hourly TPWS observations and NWP data. The state-space equations are
\begin{align}
    \label{eq:state_transition_eq}
    \h_t &= \A \h_{t-1} + \etabf_h \\
    \v_t &= \B \h_{t} + \etabf_v
\end{align}
%
%
%
%
where $\etabf_h$ and $\etabf_v$ are zero mean Gaussian distributions with covariances $\Sigmabf_h \in \mathbb{R}^{H \times H}$ and $\Sigmabf_v \in \mathbb{R}^{V \times V}$ respectively. The matrix $\A \in \mathbb{R}^{H \times H}$ is the state transition matrix and $\B \in \mathbb{R}^{V \times H}$ is the measurement matrix.

The data is assumed to comprise an offset and a diurnal periodic component with a period of $\tau=24$ hours. The DLM models this by combining a local-level and a free-form seasonal model \cite{west2006bayesian} to produce the state transition matrix
\begin{align}
    \A =
    \begin{scriptsize}
        \begin{bmatrix}
            1 &  0 &  0  & \cdots &  0 &  0 \\
            0 & -1 & -1  & \cdots & -1 & -1 \\
            0 &  1 &  0  & \cdots &  0 &  0 \\
            0 &  0 &  1  & \cdots &  0 &  0 \\
            \vdots & \vdots &  \vdots & \ddots &  \vdots &  \vdots \\
            0 &  0  &  0 & \cdots &  1 &  0
        \end{bmatrix}
    \end{scriptsize}
\end{align}
%
Given $\tau$ and the structure of $\A$, the latent dimension is $H=24$.

The DLM naturally performs data fusion when the observed variable comprises multiple data sources measuring the same latent variable. With the TPWS and NWP data $V=2$, the observed variable is  $\v_t = \begin{bmatrix} x_t & y_t \end{bmatrix}^\top$, and
\begin{align}
    \B =
    \begin{scriptsize}
        \begin{bmatrix}
            1 & 1 & 0 \cdots & 0 \\
            1 & 1 & 0 \cdots & 0
        \end{bmatrix}
    \end{scriptsize}
\end{align}
%


$\Sigmabf_v$ represents the covariance of the data noise and determines how sensitive the model is to a particular data source. With an estimate of the data noise for the TPWS and NWP data given by $\epsilonbf = \begin{bmatrix} \epsilon_x & \epsilon_y \end{bmatrix}^\top$, 
the covariance matrix is
\begin{align}
    \Sigmabf_v = 
    \begin{bmatrix}
        \epsilon_x^2 & 0 \\
        0 & \epsilon_y^2
    \end{bmatrix}
\end{align}
We also consider a model that excludes the NWP data. For this, $V=1$ and the second dimension is removed in $\v_t$, $\B$, and $\Sigmabf_v$.

Setting the values of $\Sigmabf_h$ generally lower than $\Sigmabf_v$ assumes that the data is more uncertain than the model's dynamics contained in (\ref{eq:state_transition_eq})
\footnote{E.g. see the function of the Kalman gain (\ref{eq:kalmanGain}) in equations (\ref{eq:kalmanUpdatef}) and (\ref{eq:kalmanUpdateF})}.
We thus set
\begin{align}
    \Sigmabf_h = 0.8 \I \min(\epsilon_x^2, \epsilon_y^2)
\end{align}
where $\I$ is the $H \times H$ identity matrix.

\subsection{Noise Estimation}

The data noise $\epsilonbf$ used for the DLM model is estimated by extracting it from the data using a high pass filter. 
A $3^\text{rd}$ order Butterworth high-pass filter is used, which provides a roll off of -18 dB/decade \cite{oppenheim2010discrete}. A cut-off frequency of $5/\tau$ is used to remove low frequency components such as the diurnal cycle. The noise of each data stream is then represented by the standard deviation of the high-pass-filtered data stream and stored in $\epsilonbf = \begin{bmatrix} \epsilon_x & \epsilon_y \end{bmatrix}^\top$.

In our experiments, we found that the approach can occasionally underestimate the true noise, resulting in over confident predictions. A lower threshold is thus applied to the noise estimate. The lower threshold is 0.7 for the TPWS data and 1.5 times the TPWS noise for the NWP data. A higher NWP data noise encourages the model to rely more on the TPWS data.

%

\subsection{Kalman Filtering}

The Kalman filter performs inference on the DLM by computing the posterior filtered distribution
$p(\h_t | \v_{t-T:t}) = \mathcal{N}(\f_t, \F_t)$,
%
where $\f_t$ is the filtered mean, $\F_t$ is the filtered covariance, and $\mathcal{N}$ is the Gaussian. The Kalman filter operates in a two-step prediction-update process defined by the following equations:
\\
Prediction step:
\begin{align}
    \mubf_h &= \A \f_{t-1} \\
    \Sigmabf_{hh} &= \A \F_{t-1} \A^\top + \Sigmabf_h \\
    \mubf_v &= \B \mubf_h \\
    \Sigmabf_{vv} &= \B \Sigmabf_{hh} \B^\top + \Sigmabf_v
\end{align}
Update step:
\begin{align}
    \label{eq:kalmanGain}
    \K &= \Sigmabf_{hh} \B^\top \Sigmabf_{vv}^{-1} \\
    \label{eq:kalmanUpdatef}
    \f_t &= \mubf_h + \K ( \v_t - \mubf_v ) \\
    \label{eq:kalmanUpdateF}
    \F_t &= (\I - \K \B)\Sigmabf_{hh}
\end{align}

The computed $\f_t$ and $\F_t$ can be transformed back to the observation space to provide an estimate of the variable of interest in the form $\mathcal{N}(\mu_t, \sigma_t^2)$ using
\begin{align}
    \mu_t &= \C \f_t \\
    \sigma_t^2 &= \C \F_t \C^\top + \epsilon_x
\end{align}
%
%
%
where $\C = \begin{bmatrix} 1 & 1 & 0 & \dots & 0 \end{bmatrix}$, which is identical to the first row of $\B$.

At time $t=1$, the Kalman filter update equation requires $\f_0$ and $\F_0$. We set $\F_0 = 5\Sigmabf_h$ and $\f_0$ is set using a sequence of $T$ calibration data samples. Assuming $T$ is divisible by $\tau$, the $i^\text{th}$ element of $\f_0$ is set as
\begin{align}
    f_0[i] =
    \begin{cases}
        \frac{1}{T} \sum_{t=1}^T x_t & i=0 \\
        \frac{\tau}{T} \sum_{t=1}^{T/\tau} x_{\tau (t-1)+i} & i = 1, 2, \dots H-1
    \end{cases}
\end{align}
%

\subsection{Evaluation}

A test sample $x_t$ can be evaluated according to the predicted distribution $\mathcal{N}(\mu_t, \sigma_t^2)$ using to the two-sided $p$-value 
\begin{align}
    p_\text{two-sided}(x_t | \mu_t, \sigma_t^2) = 1 - 2 \left| 0.5 - \Phi\left( \frac{x_t - \mu_t}{\sigma_t} \right) \right|
\end{align}
where $\Phi$ is the cumulative distribution function. The test sample $x_t$ is suspect if
\begin{align}
    \label{eq:pTest}
    p_\text{two-sided}(x_t | \mu_t, \sigma_t^2) < 0.1
\end{align}
%

%

\section{Datasets and Testing}

\begin{table}[!t]
    \centering
    \caption{Dataset properties. ($\mathcal{U}$ denotes a uniform distribution).}
    \label{table:datasets}
    \setlength{\tabcolsep}{5pt}
    \begin{scriptsize}
        \begin{tabular}{lccc}
            \toprule
            Dataset & No. of stations & Period & Test Samples \\
            \midrule
            Temperature &  99 	& 2017 - 2018 & daily minimum \\
            Humidity 	& 100 	& 2018 - 2019 & 9am and 3pm \\
            Wind gust 	& 100 	& 2016 - 2020 & daily maximum \\
            \bottomrule
        \end{tabular}
    \end{scriptsize}
\end{table}

In this study data from BoM weather stations are used as TPWS data and the BoM's ACCESS-R \cite{bom2010operational} provides the NWP data source. We consider daily minimum temperatures, daily maximum wind gusts, and relative humidity observations at 9am and 3pm. Data are gathered for $\sim$100 stations across Australia, each with at least two years of data. See Table \ref{table:datasets}. The ACCESS-R data are only available for temperature, and thus the DLM for wind gusts and relative humidity considers the TPWS data streams only.

To test the approach, synthetic datasets are created where data samples are randomly perturbed. A total of 2.7\% of minimum temperatures perturbed with uniform distributions $\mathcal{U}(-6^\circ C, -2^\circ C)$ or $\mathcal{U}(2^\circ C, 6^\circ C)$. Similarly, a total of 10\% of maximum wind gusts perturbed with $\mathcal{U}(5 m/s, 14.6m/s)$.

Directly perturbing relative humidity can violate the relative humidity range of [0\%, 100\%]. Thus, the dew point temperature is perturbed with $\mathcal{U}(-10^\circ C, -4^\circ C)$ or $\mathcal{U}(4^\circ C, 10^\circ C)$ and translated to relative humidity according to \cite{bom2009calculation}:
\begin{align}
    \label{eq:rhConversion}
    \text{RH} = \frac{100 \exp \left( 1.8096 + \frac{17.2694 \text{T}_d}{237.3 + \text{T}_d} \right)}{ \exp \left( 1.8096 + \frac{17.2694 \text{T}_a}{237.3 + \text{T}_a} \right) }
\end{align}
Here $\text{T}_d$ is the dew point temperature and $\text{T}_a$ is the dry bulb temperature. Similarly, the DLM is also applied to dew point temperature and the results are converted to relative humidity via (\ref{eq:rhConversion}).

During testing, the DLM is applied to the perturbed hourly data-streams for each TPWS and each weather variable. The perturbed test samples are compared with their corresponding DLM predictions and classified as valid or suspect according to (\ref{eq:pTest}). The hit rate (number of true positives over the total number of positive cases) and false positive rate (number of false positives over the total number of negative cases) are computed from the classifications. 

\section{Results}

The average hit rate, false positive rate, and accuracy for the various datasets are presented in Table \ref{table:results}. The high hit rates indicate that the model is able to detect the majority of the perturbed samples. The low false positive rates indicate a low probability of incorrectly classifying a valid sample as a suspect. 
\begin{table}[!t]
    \centering
    \caption{Average performance results.}
    \label{table:results}
    \setlength{\tabcolsep}{10pt}
    \begin{scriptsize}
        \begin{tabular}{lccc}
            \toprule
            Dataset 		& Hit rate (recall)	& False positive rate	& Accuracy \\
            \midrule
            Temperature		& 80.0\% 			& 10.3\% 				& 88.9\% \\
            Humidity		& 91.4\%			& 7.78\%				& 92.2\% \\
            Wind Gust		& 84.0\%			& 8.51\%				& 91.0\% \\
            \bottomrule
        \end{tabular}
    \end{scriptsize}
\end{table}

\begin{figure}[!t]
    \centering
    \includegraphics[width=\columnwidth]{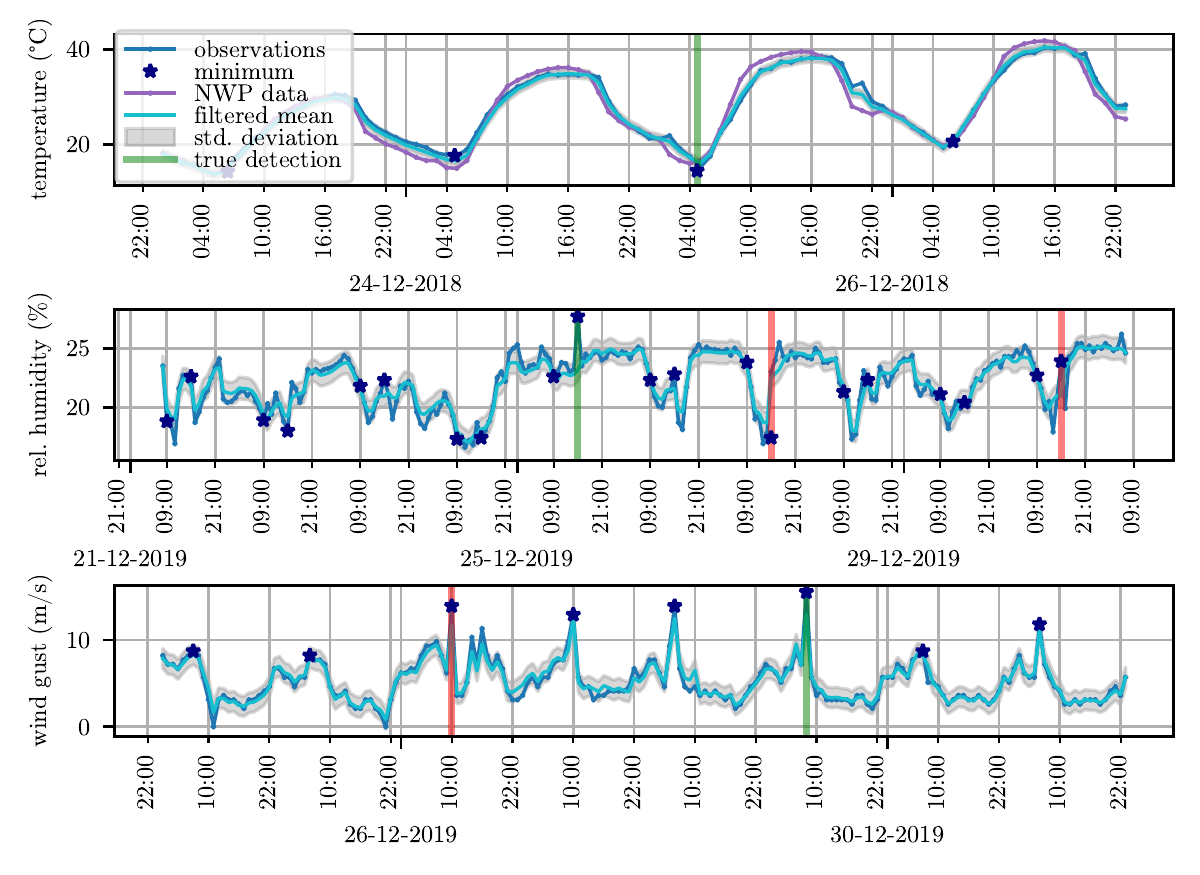}
    \caption{Plots with examples of true and false detections.}
    \label{fig:allPlots}
\end{figure}

A plot of a true detection for temperature is illustrated in the top panel of Fig. \ref{fig:allPlots}. The perturbed value is not significantly lower than the predicted minimum, however it is still detected. This particular TPWS provides a relatively smooth diurnal cycle and this has detected through the noise estimation. As such, the DLM is confident in the data and produces a low variance in its predictions. Small perturbations are thus able to be detected.

\begin{figure}[!t]
    \centering
    \includegraphics[width=\columnwidth]{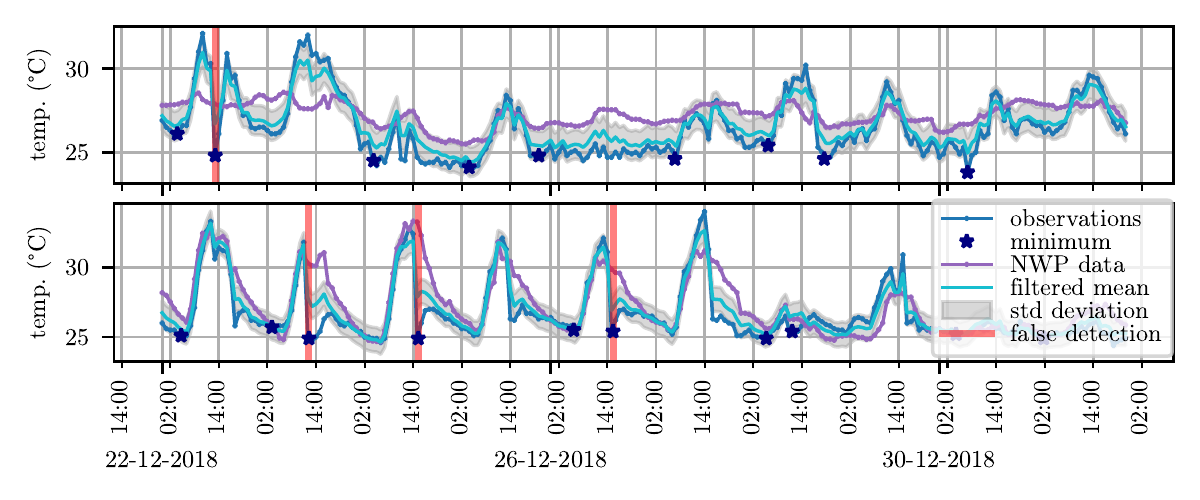}
    \caption{Examples of conditions that cause false or missed detections.}
    \label{fig:failurePlots}
\end{figure}

Not all TPWS produce smooth diurnal cycles. In general, the approach can fail when there is (1) a significant difference between the TPWS and the NWP data, (2) a lack of a diurnal cycle, or (3) a severe non-linear change in the dynamics. Examples of this are illustrated in Fig. \ref{fig:failurePlots}. In the top panel the NWP data differs from the TPWS observations and there is also a poor diurnal cycle in the data. The result is that a high noise level is estimated and the model predictions are associated with high variance. The model is thus desensitized to noise, reducing its ability to detect the perturbations. In the bottom panel of Fig. \ref{fig:failurePlots}, the NWP underestimates the minimum temperatures and the TPWS data contain non-linear changes in the dynamics, resulting in false positive detections.

Plots of true and false detections for relative humidity and wind gust are presented in the middle and bottom panels of Fig. \ref{fig:allPlots} respectively. These variables generally do not have as strong diurnal cycles as temperature and the data has a higher noise levels. The DLM however generally performs better on these datasets. This is mostly due to the model not using the NWP data. Using NWP data assists with identifying errors such as biases, but it can also have a negative affect when its predictions are not accurate. This usually occurs when the NWP is a low resolution prediction and is not designed to predict highly localised weather affects. Selecting an accurate and high resolution NWP model is crucial.

%


\section{Conclusion}

We present an approach to detect unusual readings from Third-Party Weather Station data and test it on daily minimum temperature, maximum wind gust, and relative humidity. We show that the model is robust in that the same model parameters function over a range of weather stations and weather variables.

Our approach involves Kalman filtering and an initial model calibration via noise estimation. The approach is simple to implement and maintain, making it well suited to production environments. Finally, given that the approach is a time-series based approach, we believe that it will complement the common statistical and heuristic approaches typically used for weather data quality control.

\section*{Acknowledgement}

We thank the BoM for their assistance in providing the data (including the perturbed data) and for the many discussions on quality control.

\bibliographystyle{unsrtnat}
\bibliography{bibliography}

\end{document}